\documentstyle[prl,aps]{revtex}
\tighten
\begin{document}
\draft
\twocolumn
\title{Bures distance between two displaced thermal states }

\author{Gh. - S. Paraoanu\thanks{On leave of absence from the Department of Theoretical Physics, Institute of Atomic Physics, PO Box MG - 6, R - 76900, Bucharest--Magurele, Romania.}}

\address{Department of Physics, University of Illinois at Urbana--Champaign, 1110 W. Green St., Urbana, IL 61801, USA; e-mail: paraoanu@physics.uiuc.edu}

\author{Horia Scutaru }

\address{Department of Theoretical Physics, Institute of
Atomic Physics, PO BOX MG-6,
 R-76900 Bucharest-Magurele, Romania; e-mail: scutaru@theor1.ifa.ro}


\maketitle

\begin{abstract}
The Bures distance between two displaced thermal states and the corresponding geometric quantities (statistical metric, volume element, scalar curvature) are computed. Under nonunitary (dissipative) dynamics, the statistical distance shows the same genera

l features previously reported in the literature by Braunstein and Milburn for two--state systems. The scalar curvature turns out to have new interesting properties when compared to the curvature associated with squeezed thermal states. 
\end{abstract}


In the geometry of mixed (impure) quantum states few concrete
results have been found. The Bures distance and the metric
have been computed for the spin-${1 \over 2}$ system [1] and
for the spin-1 system [2]. Recently the Bures distance between two
undisplaced thermal squeezed states was obtained [3]. This is a 
remarkable result because it is the first example 
of this type in an infinite
dimensional Hilbert space. A general formula for the transition
probability between any impure state and a pure squeezed state
was obtained in [4]. 

A class of thermal states which are not treated in [3] is that of
displaced thermal states, also called coherent thermal states [5].

The main results of this paper are  the formulae for the transition
probability and the Bures distance between two displaced thermal states. 

The transition probability between two quantum states described
by the density matrices $\rho_{1}$ and $\rho_{2}$ in the Hilbert
space ${\cal H}$ is given by [6] 
\begin{equation} 
P(\rho_{1},\rho_{2}) = \left({\rm Tr}
\sqrt{\sqrt{\rho_{1}}\rho_{2}                 
\sqrt{\rho_{1}}} \right)^2, 
\end{equation}
and the Bures distance is given by
\begin{equation} 
D_{B}^2(\rho_{1},\rho_{2}) = 2 \left [1 - \sqrt{P(\rho_{1},
\rho_{2})} \right]. \label{bur}
\end{equation}  
The density matrix of a displaced thermal state is parametrized
in the form 
\begin{equation}
\rho(\beta,(p,q)) = D(p,q)\rho(\beta)D(p,q)^{\dagger}, 
\end{equation}
where $\rho(\beta) = \frac{1}{Z(\beta)}\exp(-\beta H)$  with  
\newline $H={1 \over 2}(P^2+Q^2)$,  $[Q,P]=iI$,
\newline $Z(\beta) = {\rm Tr} \exp(-\beta H)=
(2 \sinh \frac{\beta}{2})^{-1}$ and $D(p,q) = \exp i(pQ-qP)$. Here $p$ and $q$ are the displacements in momentum and coordinate and $\beta$ is the inverse temperature. 
Then, as it follows from [3], one has $\sqrt{\rho(\beta,(p,q))}=
D(p,q)\sqrt{\rho(\beta)}D(p,q)^{\dagger}$. It is well known [6]
that the transition probability is invariant under the unitary
transformations in ${\cal H}$
\begin{equation}
P(U\rho_{1}U^{\dagger}, U\rho_{2}U^{\dagger}) =
P(\rho_{1}, \rho_{2}), 
\end{equation}
and is symmetric
\begin{equation} 
P(\rho_{1}, \rho_{2}) = P(\rho_{2}, \rho_{1}).
\end{equation}
Hence it suffices to compute ${\rm Tr} \sqrt{A^{\dagger}A}$
where 
\newline $A = \sqrt{\rho(\beta_{2},(p,q))}\sqrt{\rho(\beta_{1})}$
and $p=p_{2}-p_{1}$, $q=q_{2}-q_{1}$.
The following equation, proven in [7], allows the computation
of $\sqrt{A^{\dagger}A}$
\begin{eqnarray}
&&
\exp\{\gamma_{1}[(P-\nu_{1})^2+(Q-\omega_{1})^2]\}
\nonumber \\
&&
\exp\{\gamma_{2}[(P-\nu_{2})^2+(Q-\omega_{2})^2]\} = 
\nonumber \\
&& 
\exp\{(\gamma_{1}+\gamma_{2})[(P-\nu)^2+(Q-\omega)^2]+\theta I\},
\end{eqnarray}
with 
\newline $\theta = [(\nu_{1}-\nu_{2})^2+(\omega_{1}-\omega_{2})^2]
f(\gamma_{1},\gamma_{2})$, 
\newline $\nu = \nu_{1}g(\gamma_{1},\gamma_{2})+
\nu_{2}g(\gamma_{2},\gamma_{1}) - i(\omega_{2}-\omega_{1})f(\gamma_{1},
\gamma_{2})$, 
\newline and 
\newline $\omega= \omega_{1}g(\gamma_{1},\gamma_{2})+
\omega_{2}g(\gamma_{2},\gamma_{1}) - i(\nu_{1}-\nu_{2})f(\gamma_{1},
\gamma_{2}).$ 
The following notations have been used 
\begin{equation} 
f(\gamma_{1},\gamma_{2}) ={ \sinh\gamma_{1}\sinh\gamma_{2}
\over \sinh(\gamma_{1}+\gamma_{2}) }, ~~
g(\gamma_{1},\gamma_{2}) ={ \sinh\gamma_{1}\cosh\gamma_{2}
\over \sinh(\gamma_{1}+\gamma_{2}) }.
\end{equation}
Then
\begin{eqnarray}
&&
A = \frac{1}{\sqrt{Z(\beta_{1})Z(\beta_{2})}}
\nonumber \\
&& \exp\{-{(\beta_{1} +
\beta_{2}) \over 4}[(P-\xi)^2+(Q-\eta)^2]+\tau I\}, 
\end{eqnarray}
with 
$\tau = -(p^2+q^2)f({\beta_{1} \over 4},{\beta_{2}
\over 4})$, $\xi =pg({\beta_{2} \over 4},{\beta_{1}
\over 4}) -iqf({\beta_{1} \over 4},{\beta_{2}
\over 4})$ and $\eta= qg({\beta_{2} \over 4},{\beta_{1}
\over 4}) +ipf({\beta_{1} \over 4},{\beta_{2}
\over 4}).$
It follows that
\begin{eqnarray} 
&&
A^{\dagger} =  \frac{1}{\sqrt{Z(\beta_{1})Z(\beta_{2})}}
\nonumber \\
&&
\exp\{-{(\beta_{1} +
\beta_{2}) \over 4}[(P- \bar \xi)^2+(Q- \bar \eta)^2]+\tau I\},
\end{eqnarray}
and
\begin{eqnarray} 
&&
A^{\dagger}A  =  \frac{1}{Z(\beta_{1})Z(\beta_{2})}
\nonumber \\
&&
\exp\{-{(\beta_{1} +
\beta_{2}) \over 2}[(P- \tilde p)^2+(Q- \tilde q)^2]+(2\tau +
\tilde \tau) I\}
\end{eqnarray}
where 
\begin{equation}
\tilde \tau = 4(p^2+q^2){ (\sinh{\beta_{1} \over 4}
\sinh{\beta_{2} \over 4})^2 \over \sinh{(\beta_{1}+\beta_{2})
\over 2}}.
\end{equation}
Because
\begin{eqnarray}
&& 
A^{\dagger}A = \frac{1}{Z(\beta_{1})Z(\beta_{2})}\exp(2\tau +
\tilde \tau)
\nonumber \\
&&
D(\tilde p,\tilde q)exp\{-{(\beta_{1} +
\beta_{2}) \over 2}[P^2+Q^2] \}D(\tilde p,\tilde q)^{\dagger},
\end{eqnarray}
it follows that
\begin{eqnarray}
&& 
\sqrt{A^{\dagger}A} =\frac{1}{\sqrt{Z(\beta_{1})Z(\beta_{2})}}
\exp(\tau +
\tilde {\tau \over 2})
\nonumber \\
&&
D(\tilde p,\tilde q)\exp\{-{(\beta_{1} +
\beta_{2}) \over 4}[P^2+Q^2] \}D(\tilde p,\tilde q)^{\dagger},
\end{eqnarray}
and 
\begin{equation}
{\rm Tr} \sqrt{A^{\dagger}A} = {Z({\beta_{1}+\beta_{2} \over 2})\over \sqrt{Z(\beta_{1})Z(\beta_{2})}}
\exp(\tau + \tilde {\tau \over 2}). 
\end{equation}
The main result of the paper is
\begin{eqnarray} 
&&
P(\rho(\beta_{1}),\rho(\beta_{2},(p,q)))=
\nonumber \\
&&
\frac{Z({\beta_{1}+\beta_{2} \over 2})^2}{
Z(\beta_{1})Z(\beta_{2})}
\exp\{-\frac{Z(\beta_{1}+\beta_{2})}{
2 Z(\beta_{1})Z(\beta_{2})}
(p^2+q^2)\}.
\end{eqnarray}
Now, using (\ref{bur}), the statistical distance between any two displaced thermal states can be readily obtained. 

When $\beta_{2} \rightarrow \infty$ the state $\rho(\beta_{2},
(p,q))$ becomes a coherent state and one reobtains the well known
result [8, (4.4.15)]. When both $\beta_{1}, \beta_{2} 
\rightarrow \infty $ one obtains also the correct result [8, 
(4.4.7)].

The Bures or statistical distance metric is obtained either by considering two states close to each other and making a Taylor expansion with respect to the infinitezimal parameters or, equivalently, as [3]
\begin{eqnarray} 
ds_{B}^2 &=& g_{\mu\nu}dx^\mu dx^\nu \\
 &=& {1 \over 2} \frac{d^2}{dt^2} D_{B}^2(\rho(\beta),
\rho(\beta+t\delta \beta, t\delta p,t \delta q))|_{t=0}, 
\end{eqnarray}
which becomes in our case 
\begin{equation} 
ds_{B}^2 = \frac{1}{2}\tanh{\beta \over 2}(dp^2 + dq^2) +
{1 \over 16(\sinh{\beta \over 2})^2}d\beta^2. \label{uu}
\end{equation}

What this formula shows is that the square of the infinitezimal Bures distance consists of two parts: one given by the difference in the displacements  $p$ and $q$ and the other of thermal origin. For $\beta\rightarrow\infty$ (pure states) we recover the

Fubini-Study (Euclidean) metric $ds_{FS}^2 = \frac{1}{2}(dp^2 +dq^2)$ [9, (2.27)] for coherent states. The thermal part, $\frac{1}{16}
(\sinh\frac{\beta}{2})^{-2}d\beta^2$, which is obtained from (\ref{uu}) for $dp=dq=0$, also appears in Twamley's formula [3, (29)] as a squeezing--independent, purely thermal contribution.

Under unitary transformations, the temperature $\beta$ remains constant so only the first term of the RHS of (\ref{uu}) is responsible for changes in the statistical distance. But the temperature can be altered via non-unitary transformations. For example

, a well-known model from quantum optics [10] for the damped quantum oscillator yields the following master equation of the statistical matrix $\rho \equiv \rho (\beta , (p, q))$ 
\begin{eqnarray}
\dot{\rho}& = &- i[\omega a^{+}a, \rho ] + \gamma_{\downarrow}\{[a, \rho a^{+}]+ [a\rho , a^{+}]\} \nonumber \\
&  & + \gamma_{\uparrow}\{[a^{+}, \rho a] + [a^{+}\rho , a]\}, 
\end{eqnarray}
with $a=\frac{1}{\sqrt{2}}(Q+iP)$, $\gamma_{\downarrow}>\gamma_{\uparrow}\geq 0$, $\omega >0$. With the notations $k \equiv \gamma_{\downarrow} - \gamma_{\uparrow}$ and  $\beta_{\infty} \equiv \ln \frac{\gamma_{\downarrow}}{\gamma_{\uparrow}}$ it can be s

hown [10] that
the parameters $p$, $q$, and $\beta$, which characterize the state $\rho (\beta, (p, q))$ at any moment $t$ have the following behavior:
\begin{eqnarray}
q_{t}&=&(q_{0}\cos\omega t + p_{0}\sin\omega t)e^{-kt},\\
p_{t}&=&(-q_{0}\sin\omega t + p_{0}\cos\omega t)e^{-kt},\\
\coth\frac{\beta_{t}}{2}&=&e^{-2kt}\coth\frac{\beta_{0}}{2}+ (1-e^{-2kt})\coth\frac{\beta_{\infty}}{2}.
\end{eqnarray}
The rate of change in the statistical distance is then:
\begin{eqnarray}
\left(\frac{ds}{dt}\right)^2 &=&\frac{1}{2}\tanh\frac{\beta_{t}}{2}(k^2+\omega^2)(q_{t}^2 + p_{t}^2) \nonumber \\ & & + k^2 \sinh^2 \frac{\beta_{t}}{2}\left(\coth\frac{\beta_{t}}{2} - \coth\frac{\beta_{\infty}}{2}\right)^{2}.\label{xxii}
\end{eqnarray}
A similar quantity has been previously analyzed by Braunstein and Milburn for a two-state system under non-unitary dynamics [11]. They found that $\left( \frac{ds}{dt}\right)\rightarrow \infty$ at $t=0$ if the system is initially in a pure state.  From (\

ref{xxii}) we see that if $\beta_{0}\rightarrow\infty$ (initial pure coherent state), then indeed $\left( \frac{ds}{dt}\right)_{t=0}\rightarrow \infty$. At $t\rightarrow\infty$, the thermal contribution of $\frac{ds}{dt}$ vanishes, a feature which also ap

peared in [11]. The practical consequence of Braunstein and Milburn  results was an improvement in the accuracy of ``one-tick'' clocks; we can see now that their conclusions  can be extended from two-dimensional to infinite-dimensional spaces.  

The volume element is $dv = \frac{1}{8}{\rm sech}\frac{\beta}{2}dpdqd\beta$ and it has the significance of a quantum Jeffreys' prior; a similar quantity has been studied by Slater for finite-dimensional systems (spin $\frac{1}{2}$ and $1$) and for squeeze

d thermal states [12].

The scalar curvature associated with the Riemannian metric (\ref{uu})
is found to be
\begin{equation}
R = -6 +  14 \tanh^2{\beta \over 2}.
\end{equation} 
A new feature with respect to the result of [3] is the vanishing
of the scalar curvature for $\beta = 2\tanh^{-1}\sqrt{{3 \over 7}}$
and any value of $p$ and $q$. Also, for $\beta\rightarrow\infty$ the curvature is 8, so it does not diverge. This shows that, while $ds$ is indeed a measure of the distinguishability of two states, $R$ has a more complicated significance than that suggest

ed in [3], characterizing locally a relation between not only two, but three (or more) states. Roughly speaking, R gives the number of states equally distinguishable from a certain state $\beta , (p, q))$.
 This number depends only of temperature, and not of the other parameters of the state --- a property which was also noticed for squeezed thermal states.
 
\vfill

\bigskip

\begin{thebibliography}{99}
\bibitem{} 
M. H\"ubner,   Phys. Lett. A {\bf 163}, 239 (1992).
\bibitem{}
M. H\"ubner,   Phys Lett. A {\bf 179}, 226 (1993);

J. Dittman,   J. Geom. Phys. {\bf 13}, 203 (1994).
\bibitem{}
J.Twamley,   J. Phys. A : Math. Gen. {\bf 29}, 3723 (1996).
\bibitem{}
H. Scutaru,   Phys. Lett. A {\bf 200}, 91 (1995).
\bibitem{} 
J. Fearn  and M. J. Collet,   J. Mod. Opt. {\bf 35}, 553 (1988).
\bibitem{} 
J. Josza,   J. Mod. Opt {\bf 41}, 2315 (1994).
\bibitem{} 
R. M. Wilcox,   J. Math. Phys. {\bf 4} 962 (1967).
\bibitem{}
C. W. Gardiner,   {\it Quantum Noise} (Springer-Verlag, Berlin, 1995 ).
\bibitem{} 
J. P. Provost and G. Vallee,  Commun. Math. Phys. {\bf 76},
289 (1980).
\bibitem{}
R. S. Ingarden and A. Kossakowski,   Ann. Phys. (N. Y.) {\bf 89}, 451 (1975). 
\bibitem{}
S. L. Braunstein and G. J. Milburn,   Phys. Rev. A {\bf 51}, 1820 (1995).  
\bibitem{}
P. B. Slater,  J. Phys. A : Math. Gen. {\bf 29}, L271 (1996); L601 (1996).



\end{thebibliography}
\end{document}